\documentstyle[12pt,epsf]{article}

\begin{document}

\begin{flushright}
SLAC--PUB-7772\\
January 1999\\ 
\end{flushright}

\bigskip\bigskip
\begin{center}
{\bf\large
Some Thoughts on COMMUTATION RELATIONS and MEASUREMENT ACCURACY
\footnote{\baselineskip=12pt
Work supported by Department of Energy contract DE--AC03--76SF00515.}}

\bigskip

H. Pierre Noyes\\
Stanford Linear Accelerator Center\\
Stanford University, Stanford, CA 94309\\
\end{center}
\vfill

\begin{abstract}
We show that measuring the trajectories of charged particles to
finite accuracy leads to the commutation relations needed for
the derivation of the free space Maxwell equations using the
{\it discrete ordered calculus} (DOC). We note that the finite
step length derivation of the discrete difference version of the
single particle Dirac equation implies the discrete version
of the $p,q$ commutation relations for a free particle.
We speculate that a careful operational analysis of the
change in momenta occurring in a step-wise continuous solution
of the discrete Dirac equation could supply the missing
source-sink terms in the DOC derivation of the Maxwell equations,
and lead to a finite and discrete (``renormalized'') quantum
electrodynamics (QED).
\end{abstract}
\vfill
\begin{center}
Invited paper presented at the $19^{th}$ annual international meeting of the\\
{\bf ALTERNATIVE NATURAL PHILOSOPHY ASSOCIATION}\\
Wesley House, Cambridge, England, August 14-17, 1997\\
\end{center}
\vfill

\newpage

\section{INTRODUCTION}

Our derivation of the free space Maxwell equations using the {\it
discrete ordered calculus} (DOC)\cite{Kauffman&Noyes96a} mentioned that
the postulated commutation relations between position and velocity
could be interpreted as a consequence of a fixed discrepancy
between first measuring position and then velocity or visa versa.
However, these commutation relations were not given a careful physical
justification in terms of our finite measurement accuracy philosophy
\cite{Noyes96a}. A second deficiency, which in fact caused us to warn
the reader that we had only derived one part of the {\it formalism} of
classical electrodynamics rather than the theory itself, was that
no attempt was made to identify the sources and sinks of the ``fields''
and derive the inhomogeneous Maxwell equations from them. We took
a step in that direction by our derivation\cite{Kauffman&Noyes96b}
of a finite and discrete version of the 1+1 free space Dirac equation
from a fixed step-length {\it Zitterbewegung} postulate using finite
difference equations. Although it was noted that an attempt had been
made by me\cite{Noyes95} to attribute the {\it Zitterbewegung}
to the conservation of spin or particle number in the presence
of random electromagnetic fluctuations, no attempt was made to relate
these interactions to the source terms needed to complete the argument
in the Maxwell equations paper. Neither Kauffman nor I
have attempted to relate the
non-commutativity known to arise from the Dirac equation
to the commutation relations needed to derive the Maxwell equations
in our finite and discrete context. In this paper I take a few steps to
remedy both defects, but more work is needed.

\section{ELECTROMAGNETIC MEASUREMENT OF \break A CHARGED PARTICLE TRAJECTORY}

In earlier work I have made use of what I called ``the counter paradigm''
to cut the Gordian knot of specifying what a physicist means when he
says that a particle was or was not present in a finite spacial volume
for a finite time duration. As a first approximation, I assume that
this volume is the ``sensitive volume'' of a counter, and the time
duration is the time during which the recording device attached
to the counter could have recorded an event, often called a
``firing''. This I call a NO-YES event, depending on whether
the counter did not or did ``fire''. A more careful treatment specifies the
probability of ``spurious events'', i.e. cases when the counter ``should
have fired'' but did not (counter inefficiency), and the probability
of cases when the counter ``should not have fired'', but according to
the record did in fact fire (background events). Ted Bastin has often objected that this
abrupt transition from the laboratory to Boolean logic sweeps too
much under the rug, and I have often replied that to justify
this way of talking about laboratory practice would require a book.
Fortunately, Peter Galison has taken ten years to write the book
I needed. He separates the history of the material
culture of particle physics into a
``logic'' tradition contrasted with an ``image'' tradition.
My ``counter paradigm'' finds its appropriate niche as part of the
logic tradition. Galison shows that
by now the two alternatives have fused in the mammoth ``detectors''
which are integrated into the
accelerators in all high energy particle physics laboratories
\cite{Galison97}. It took over a century for this language and practice
to mature, and a decade to make a convincing argument as to why it
should be accepted by philosophers. I now have a simple tactic open.
I can ask any critic of my conceptual leap from counter firings to
NO-YES events to first convince me that Galison's defense of the
mainstream tradition is inadequate. Only then will I feel any need to take
his or her criticism seriously.

This ploy allows me to use conventional language in my descriptions of
laboratory measurement. In particular I can now construct
a simple paradigm for what I mean by
the measurement of the electromagnetic trajectory of a particle.
First recall that by a ``particle'' I mean\cite{Noyes96a} ``a
conceptual carrier of conserved quantum numbers between events''.
I can take the simplest interpretation of two sequential counter firings
a fixed distance $L$ apart with a time interval $T$ between them
to be that a particle conserving mass, momentum and energy passed
between them with velocity $L/T$. I assume available a  ``source''
of particles which allows a
large number of repetitions of these paired sequential events to occur.
This data set is assumed to provide both statistical and systematic
accuracy adequate for calibrating the {\it changes} in
the magnitude and/or direction of this velocity caused by
inserting electromagnetic devices into the path defined
by sequential counter firings

The electromagnetic device we consider first, inserted between
two counters previously used to measure velocity, is simply
two parallel conducting plates with a hole through them
across which a constant voltage can be applied.
This voltage is measured by standard techniques.
When the voltage is negligible, our original
source and sink counters still give a velocity $v=L/T$ for each
particle ``passing through the two holes'', showing that we
can maintain the same particulate interpretation of the two sequential
events with the plates in place, even though we do {\it not}
``measure'' the presence of the particles between the plates.
We now apply a voltage $V$ across the plates, which are large
enough compared to the holes so that, according to standard
electrostatic theory, the electric field between the plates
and along the direction of motion of the particle is ${\cal E}=V/\Delta
d$ where $\Delta d$ is the separation between the plates.
We now study the {\it change} in the velocity of
a particle of the type being studied (i.e. produced in the same way
or available from the same source)
during a time when the voltage across the plates is held at $V$.
Counter firings before the presumed arrival and after the
presumed departure of the particle at the device allow us to say
that the particle arrived at the position of
the plates with velocity $v_1$ and left with velocity $v_2$.
We then say that the particles have a charge $e$, a (rest) mass $m$,
an energy $E_1$ before they enter the first hole, and an energy $E_2$
after leaving the second hole when, for various experiments, the
velocity change produced by the device is equivalent to an energy change 
\begin{equation}
\Delta E= E_2-E_1 = \pm e{\cal E}\Delta q; \ \ \  {\cal E}=V/\Delta d
\end{equation}
with
\begin{equation}
E_1={m\over \sqrt{1 - (v_1^2/c^2)}} ; \ \ \
E_2={m\over \sqrt{1 - (v_2^2/c^2)}}
\end{equation}
We then take this as our paradigm for the {\it measurement} of an
electric field in a region of length $\Delta d$ of strength
${\cal E}$.

We emphasize that this measurement requires a {\it change} in
the velocity of the particle. The minimum change to which we can
reliably assign a number {\it quantizes} our {\it measurement accuracy}
at the level of technology we are using. Note that our
paradigm assumes {\it constant} velocity between measurements
in field-free regions. [Recall that we {\it derived} a
discrete version of the constant velocity law from
bit-string physics in our foundational paper\cite{Noyes&McGoveran89},
Sec. 6.5, pp 94-95.]
Alternatively, if we know the field
(or voltage) and the (constant velocity) trajectories before
and after the device, we can use the {\it same} device as
a paradigm for {\it position} measurement to an accuracy $\Delta d$.
By fleshing out this paradigm, we can recursively use electromagnetic
language to justify the construction of laboratory counters which
have a conceptual connection to those used in our counter paradigm.

Our paradigm for magnetic field (or momentum) measurement assumes
that we have two double plates across each of which independently adjustable
voltages can be applied. We call the entrance hole of the first pair 1
and the exit hole 2, and for the second plate the entrance hole 3 and
the exit hole 4; thus the gaps are $d_{12}$ and $d_{34}$, and the
trajectory is 1,2,3,4.
The plates are located geometrically in
the laboratory in such a way that a path connecting the exit hole 2
from the first pair to the entrance hole 3 into the other can be
an arc of a circle of radius R whose center lies in a plane with the
two gaps; the gaps between the plates are
two (short) arcs of that circle. The arc
between the two devices is  of length $R\Delta \theta$. The magnetic field we wish
to measure is perpendicular to the plane of the circle and is
of constant strength ${\cal B}$, along this arc. This is ``guaranteed''
by the geometry and the standard theory of magnetostatic fields.
According to electromagnetic
theory, this field does not change the energy of the particle,
or the magnitude of its velocity, but does cause the {\it direction}
of the velocity to change. This change is simply described in
terms of the momentum ${\bf P}$ of vector magnitude
\begin{equation}
{\bf P} = {m{\bf v}\over \sqrt{1 - (v^2/c^2)}}; \ \ \
|{\bf v}|={R\Delta \theta\over t_3-t_2}
\end{equation}
where the time $t_2$ when the particle exits hole 2 and the time $t_3$
when it enters hole 3 are usually inferred rather than directly
measured; {\bf v} is the vector velocity of constant magnitude
with a (varying) direction assumed tangent to the arc.
The radius of the circle is related to the magnitude of the
momentum by
\begin{equation}
R = {eP\over c {\cal B}}
\end{equation}
and the change in momentum (due to change in direction since the
magnitude is constant) by
\begin{equation}
\Delta P = 2P \ sin^2 \Delta \theta/2 = P(1 - cos \ \Delta \theta)
\end{equation}

As as in the case of electric field measurement, we can consider
this arrangement as either a measurement of the field ${\cal B}$
at (perpendicular to) the arc $R\Delta \theta$ geometrically defined
or as a measurement of the velocity of the particle along that arc.
But as a velocity measurement, it is important to realize that
there is an ambiguity as to whether this is the measurement of
velocity {\it after} the particle has traversed the first
double plate 12, which could be a counter measuring position,
or a measurement of velocity {\it before} it traverses the
second double plate 34.

If all we have available are not individual particle detectors,
but only devices that measure the charged current flowing along the
trajectory, the arrangement discussed above can only be used to
measure $e/m$ and not charge and mass separately. Such experiments
were, historically, sufficient to convince the proponents of various
models of the charge distribution ``within the electron'' (Abraham,
Lorentz, Poincar{\'e}) that
their models were wrong, and that the Einstein equation connecting
mass to velocity used above was correct even though it violated
their way of thinking about space and time
(\cite{Galison97}, Sec. 9.6, pp 810-816).
Galison shows by this historically examined case
that experimental tradition and the material culture of physics
allow theoretical physicists on opposite sides of what Kuhn would call
a ``paradigm shift'' to agree on the significance of experimental
results..

The fact that electric and magnetic fields acting on a moving charge
effect changes in velocity along or at right angles to the
direction of motion respectively allows one to build
a ``velocity selector'' by setting up a region of electrostatic
and magnetostatic fields in which the fields are at
right angles to each other and both are at right angles to
the direction of motion of the charge. The force on the
charge due to the electric field is $e{\cal E}$ while the force
due to the magnetic field is $ev{\cal B}/c$ and the geometry
we have specified requires these forces to be in the same direction.
Consequently there is a unique direction for which they cancel,
provided the velocity has magnitude $v = c{\cal E}/{\cal B}$.
A particle of that charge with any other velocity will be
deflected away from this direction.

At first glance, such a device would seem to allow us to measure
position and velocity ``simultaneously''. But this is not correct.
So long as the charged particle has this velocity and the magnitude
and direction of the fields does not change along this straight
line trajectory, no force acts and the particle maintains constant
velocity. However, we have no way of knowing {\it where} it is within
this region, and hence when it enters and leaves it, without
a measurement. But this measurement will change the velocity. So we must measure
when the particle enters the region {\it and} when it leaves the
region in order to know how long and when it is in the region
with that velocity. As before, we can first measure position and then
velocity or first measure velocity, and then position but not
both simultaneously. An extended discussion of this case should allow
us to see that three points on the trajectory are needed to establish
the field at the intermediate point, and four if we are to measure both
${\cal {\bf E}}$ and ${\cal {\bf B}}$. On another occasion we hope to
be able to go on to derive the free field commutation relations
by such considerations (or directly from our DOC equations), and not
just the uncertainty principle restrictions obtained by Bohr and
Rosenfeld.

In closing we note that, even though we started out to
devise a paradigm for electromagnetic field measurement,
we have ended up deriving from this paradigm
the DOC postulate that we can first measure position and
then velocity or first measure velocity and then position,
but not both simultaneously.
We hope that this discussion makes it less of a mystery why the DOC
postulate leads so directly to the formalism of free-field
electromagnetism.

\section{FROM FREE DIRAC PARTICLES \break TO FIELD SOURCES AND SINKS}

The derivation of the finite difference version of the free particle
Dirac equation\cite{Kauffman&Noyes96b} for fixed step length
$\hbar/mc$ with step velocity $\pm c$ tells us immediately
that we can cut the trajectory of a free particle into
segments of constant velocity between ``points'' at which
the velocity changes discontinuously. On the other hand
our DOC equations for the free space electromagnetic field
\cite{Kauffman&Noyes96a} support solutions corresponding to the
propagation of crossed electromagnetic fields with velocity $c$
and constant frequency which, for finite segments, can be
interpreted as ``photons'' if they have the right amplitude.
All we seem to need to produce a quantum electrodynamics
which is finite and discrete, and hence ``born renormalized'',
would seem to be to assign a charge to the massive particle
which satisfies the Dirac equation in such a way that
its discrete changes in velocity correspond to the emission
or absorption of such photons. I hope to do this on another occasion.
The details will obviously take some time to work out, but will
provide a lot of fun along the way.

Since this amounts to solving a finite and discrete ``three particle
problem'', an approach to the same theory which starts more
directly from bit-string physics would be to treat the photon
as a bound state of a particle-antiparticle pair in the
relativistic three body theory now under active development
\cite{Noyes&Jones98}.


\footnotesize

\begin{thebibliography}{99}

\bibitem{Galison97}
P.Galison, {\it Image and Logic: A Material Culture of Microphysics},
Chicago University Press, 1997.

\bibitem{Kauffman&Noyes96a}
L.H.Kauffman and H.P.Noyes, {\it Proc. R. Soc. Lond.}, {\bf A 452},
81-95 (1996).

\bibitem{Kauffman&Noyes96b}
L.H.Kauffman and H.P.Noyes, {\it Physics Letters}, {\bf A 218}, 139-146
(1996).

\bibitem{Noyes95}
H.P.Noyes, {\it Physics Essays} {\bf 8}, 434 (1995).

\bibitem{Noyes96a}
H.P.Noyes, {\it Science Philosophy Interface}, {\bf 1}, 54-79 (1996).

\bibitem{Noyes&McGoveran89}
H.P.Noyes and D.O.McGoveran, {\it Physics Essays}, {\bf 2}, 76-100
(1989).

\bibitem{Noyes&Jones98}
H.P.Noyes and E.D.Jones, ``Solution of a Relativistic Three Body
Problem'', submitted to {\it Few Body Systems}, and SLAC-PUB-7609,
(rev. June 1998).
\end{thebibliography}
\end{document}